\documentclass[final,3p,times,twocolumn]{elsarticle}
 \usepackage{graphics}
\usepackage{amssymb}
\usepackage{amsmath}

\usepackage{lineno}

\usepackage{subfigure}
\usepackage{color}
\definecolor{dgreen}{rgb}{0,0.8,0}
\definecolor{violet}{rgb}{0.5,0,0.5}

\journal{Nuclear Instruments and Methods A }

\begin{document}

\begin{frontmatter}



%% SINGLE AUTHOR
\title{Simulation of beam induced lattice defects of diamond detectors using FLUKA}

\author[kit,focal]{Moritz Guthoff\corref{cor1}}
\ead{moritz.guthoff@cern.ch}
\author[kit]{Wim de Boer}
\author[kit,focal]{Steffen M\"uller}

\cortext[cor1]{Principal Corresponding Author}
\address[kit]{Institut f\"ur Experimentelle Kernphysik, Karlsruhe Institute of Technology, Campus S\"ud, P.O. Box 6980, 76128 Karlsruhe, Germany}
\address[focal]{CERN, CH-1211 Geneva 23, Switzerland}

\begin{abstract}

Diamond is more and more used as detector material for particle detection. One argument for diamond is its higher radiation hardness compared to silicon. Since various particles have different potential for radiation damage at different energies a scaling rule is necessary for the prediction of radiation damage. For silicon detectors the non-ionising energy loss (NIEL) is used for scaling the effects of different particles. A different way of predicting the radiation damage is based on the Norget-Robinson-Torrens theorem to predict the number of displacements per atom (DPA). This provides a better scaling rule since recombination effects are taken into account. This model is implemented in the FLUKA Monte Carlo simulations package for protons, neutrons and pions. We compare simulation results of NIEL and DPA for diamond and silicon material exposed to protons, neutrons and pions for a wide range of energies.

\end{abstract}


\end{frontmatter}
% \linenumbers

\section{Introduction}
The radiation hardness of diamond as a detector material is a key argument for many of its applications. Since diamond is considered as a possible material for the upgrade of the inner tracking detectors in ATLAS and CMS \cite{ibl, cmsdiapx} it is vital to have a scaling rule for mixed field irradiation scenarios in order to predict the detector efficiency over time. The goal is to determine the damage potential of various particle types relative to others, so that the overall damage potential of a given particle energy spectrum can be predicted. In the silicon detector community the so-called NIEL-scaling hypothesis (based on non-ionising energy loss) was proven useful for most particle spectra \cite{rosetn}. This hypothesis states that the detector efficiency decreases with the number of induced lattice displacements. We studied protons, neutrons and pions with a kinetic energy between 1\,MeV and 100\,GeV, i.e. typical ranges for modern high energy experiments. This simulation is an update to a previously published paper \cite{boer07}, where only the NIEL hypothesis was considered. This paper uses a new FLUKA code \cite{Ferrari:2005zk,Battistoni:2007zzb} to determine the number of atom dislocations (DPA) caused by impinging particles. While the NIEL includes all non-ionizing energy loss, the DPA omits phonon interactions and takes recombination into account.

\section{Diamond detectors}

\subsection{Effects Influencing the detector efficiency}
\label{sec:detect_efficiency}
Many effects contribute to the signal height of a detector, of which the lattice damage is only one. Some of these effects are the following:

\begin{itemize}
\item{Bulk damage:} This is the intrinsic lattice damage caused by impinging particles. If the recoil energy is higher than the lattice binding energy, an atom is displaced from its site. These vacancies act as traps for electrons generated by ionisation and reduce their effective drift length, thus the measured signal.

\item{Polarisation effect:} The bulk damage caused by impinging particles lead to a higher trapping probability of charge carriers in the bulk material. This can lead to space charges forming an internal electric field, which is weakening the externally applied field, thus reducing the drift velocity of the charge carriers. A similar polarisation effect may occur, if charge is trapped at the interface between the diamond and the metal contact in case of a non-ohmic contact. This leads to a reduced efficiency of the detector. More details can be found in \cite{pom2008,guthoff2013}.
\end{itemize}

The detector damage is often measured indirectly by its signal yield for standardised particle types, e.g. minimum ionising particles (MIPs). An accepted figure of merit of the diamond detector quality within the diamond community is the charge collection distance CCD, which is the average drift length of the charge carriers.
To allow best possible comparison between different data sets of irradiations, a universal parameterisation, as used by the RD42 group \cite{rd42status}, is employed to determine the damage potential of different particle types: 

\begin{equation}{\textrm{CCD}(\phi)=\frac{\textrm{CCD}_0}{1+k\phi \textrm{CCD}_0},}\label{lab:ccd}\end{equation}
where CCD$(\phi)$ is the charge collection distance of the detector as a function of fluence $\phi$, $k$ is a radiation damage factor, which is dependent on the particle type and energy. CCD$_0$ the initial charge collection distance of the unirradiated diamond. The formula is motivated by the assumption that the number of traps increases with integrated luminosity and the signal is inversely proportional to the number of traps.

The more radiation damage is done by the particles, the higher is their corresponding $k$ value. By taking the ratio of $k$ for two different particle types, one can derive how many particles are needed by one type to obtain the same damage as for another particle type. For predictions of radiation damage a theoretical model is needed to describe these scaling of particle types and energies. A typical value for 24~GeV protons was measured to be $k \approx 1 \times 10^{-18}\frac{cm^2}{\mu m}$ \cite{rd42status}.

\section{Simulation of Beam Induced Lattice Defects}\label{lab:dpa}

\begin{figure*}
\begin{center}
\includegraphics[width=0.6\linewidth]{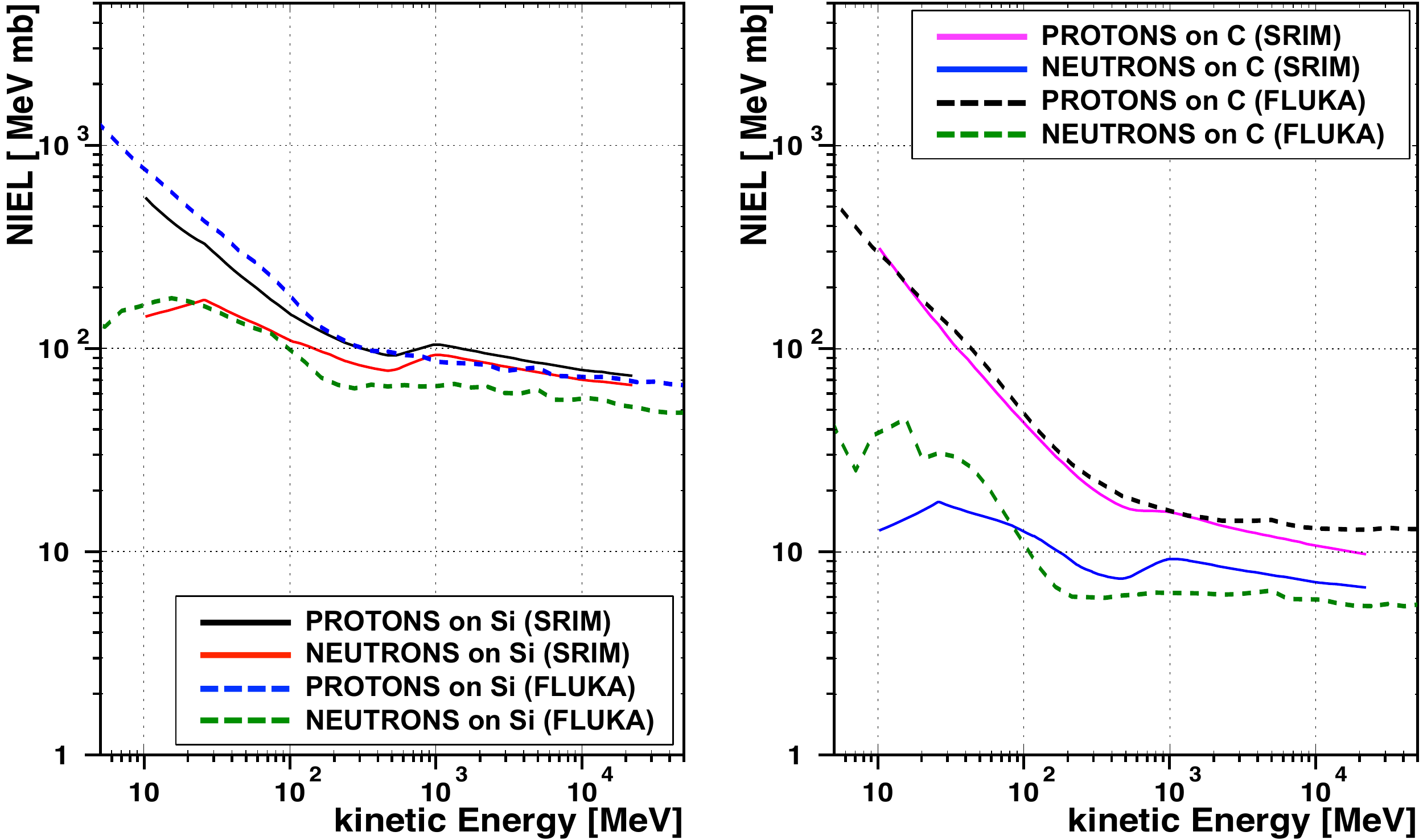}
\end{center}
\caption{NIEL cross sections for proton and neutron interactions in silicon and diamond calculated with FLUKA in comparison with a previously published plot containing results from a SRIM simulation \cite{boer07}.}
\label{fig:niel_srimvsfluka}
\end{figure*}

\subsection{Simulation Method}
To simulate the NIEL and the DPA for diamond and silicon, a simple geometry was set up. The material under test is assigned to a small box with the dimensions of $10\times 10\times 0.4$~mm$^3$, which are standard dimensions of diamond detectors installed in HEP experiments \cite{cmsbrm}. Using these dimensions the simulation result is already normalised per incoming particle per cm$^2$. The thickness is also set to standard sensor dimensions, so that secondary particle showers inside the detector are taken into account like in the real sensor. The particle beam hits the diamond volume perpendicular to the area norm, which represents the standard case for diamond detectors. However, scattered particles might increase the radiation damage in real world experiments, which needs to be taken into account, in order to accurately predict the radiation damage. 
All simulations were run, until the statistical error was at least below 5\% for all derived quantities.

\subsection{Non Ionising Energy Loss - NIEL}
While ionizing energy loss is the dominant contribution of the energy deposition from charged particles, it does not lead to significant radiation damage in silicon or diamond detectors. The non ionising energy loss (NIEL) is associated with nuclear interactions, which causes lattice defects and hence radiation damage.

The stopping power $S = dE/dx$ of a material can be expressed as the sum of energy loss transferred to electrons (ionising, $S_i$) and energy transferred to nuclei (non-ionising, $S_n$):

\begin{equation}
S = S_n + S_i = \frac{dE_n}{dx} + \frac{dE_i}{dx}
\end{equation}
The energy dependent Lindhard partition function $\xi(T)$, which goes directly into the calculation of the NIEL, is given by fraction of the total stopping power $S(T)$ that goes into the non-ionising part:

\begin{equation}
\xi(T) = \frac{S_n}{S}
\end{equation}

The NIEL is usually expressed as stopping power in units of $keV\,cm^2/g$ or as NIEL cross section in units of $MeV mb$.
In FLUKA the NIEL stopping power is not simulated, but rather the deposition of energy due to non-ionising interactions in units of $GeV$. To convert to the NIEL cross section in units of $MeV mb$ the following formula was used:

\begin{equation}
\begin{split}
\sigma_{NIEL} [MeVmb] =  \frac{NIEL[GeV] \cdot u[g/mol]}{d[cm] \cdot \varrho [g/cm^3] \cdot N_A [mol^{-1}]}\\
\cdot 1000 [MeV/GeV] \cdot 10^{27}[mb/cm^2]
\end{split}
\end{equation}
where $NIEL$ is the FLUKA simulated number, $u$ is the molar mass, $N_A$ the Avogadro constant, $d$ the thickness of the detector and $\varrho$ the density of the material.

In a previously published paper \cite{boer07} the NIEL cross\nobreakdash-sections for silicon and diamond were published. For this calculation the SRIM simulation package was used. Figure \ref{fig:niel_srimvsfluka} shows the SRIM result overlaid with the new FLUKA simulated result. There is a qualitative agreement between both results since they are in the same order of magnitude and the shapes of the curves are similar. The difference is up to factor two, visible with low energetic neutrons in the diamond case. Most likely it is due to the inelastic hadronic interactions and the production of secondaries being handled by two different models.

%One can't expect the curves obtained with two different Monte Carlo tools being exactly on top of each other. The overall agreement is satisfactory.

\subsection{Displacements Per Atom - DPA}
Displacements per atom is a measure for the radiation damage of irradiated materials. It states how often, on average, an atom in the material was displaced due to impacting particles. For example a DPA of $1 \times 10^{-22}$ means that one atom of a sample of $10^{22}$ atoms was displaced from its lattice site. The DPA value is directly related to the number of created Frenkel pairs, which are crystallographic defects, where interstitial atoms are located near vacancies in the the lattice. 

To obtain the number of displaced atoms from the DPA value per primary one has to multiply with the number of atoms in the detector. According to Avogadro's theory the number of atoms for this detector volume is:

\begin{equation}
0.04~cm^3\times 3.52 \frac{g}{cm^3} / 12 \times \textrm{N}_A = 7.07\times 10^{21}
\end{equation}

Compared to the NIEL as a measure of radiation damage, DPA is more accurate, since it does not count phonon interactions, but only the type of interactions causing a lattice defect. In addition DPA takes recombination of defects (Frenkel pairs) into account, so that this study is considered to be an improvement to the previous NIEL study.

\subsubsection{Implementation in FLUKA}
The DPA routine was recently added into version 2011.2 of FLUKA. A brief introduction of the implementation will be given here, so that the basic mechanisms and parameters are introduced. More detailed information can be found in \cite{fas10}.

To calculate the number of Frenkel pairs, FLUKA uses the theory of Norget, Robinson and Torrens \cite{nrt}:

\begin{equation}{N_F =\kappa (T)\frac{\xi(T)T}{2E_{th}},}\end{equation}

where $N_{F}$ is the number of Frenkel pairs, $\kappa(T)$ the displacement efficiency, $T$ the kinetic energy of the primary knock on atom, $\xi(T)$ the Lindhard partition function and $E_{th}$ the lattice displacement threshold energy. In the following paragraphs, a short explanation will be given, how FLUKA calculates these quantities. 

\paragraph{Displacement efficiency $\kappa(T)$} This compensation factor takes several effects into account, such as forward scattering in the displacement cascade and the recombination of Frenkel pairs due to overlap of different branches in the collision cascades. The number of remaining defects has been approximated to fit molecular dynamics simulations. 

\paragraph{Lattice displacement energy $E_{th}$} This is the average displacement energy over all crystallographic directions, and measured with dedicated experiments or lattice simulations. For the studies presented here the threshold energy used for diamond is $E_{th}=43.3$~eV, which is the average over all lattice directions from \cite{koike}. Typical values for silicon are between 21 and 25~eV. In the simulation a value of 25\,eV is used, which is the FLUKA default value for silicon. Apart from being a para\-meter for the number of generated Frenkel pairs, it is also a parameter for $\xi(T)$ where it sets the lower limit for the restricted energy loss.

\section{Results}
\label{sec:results}

\begin{figure*}
\subfigure[\textcolor{red}{NIEL} of \textcolor{blue}{Silicon}]{
\includegraphics[trim=0cm 0cm 0.5cm 0cm, clip=true,width=0.33\textwidth]{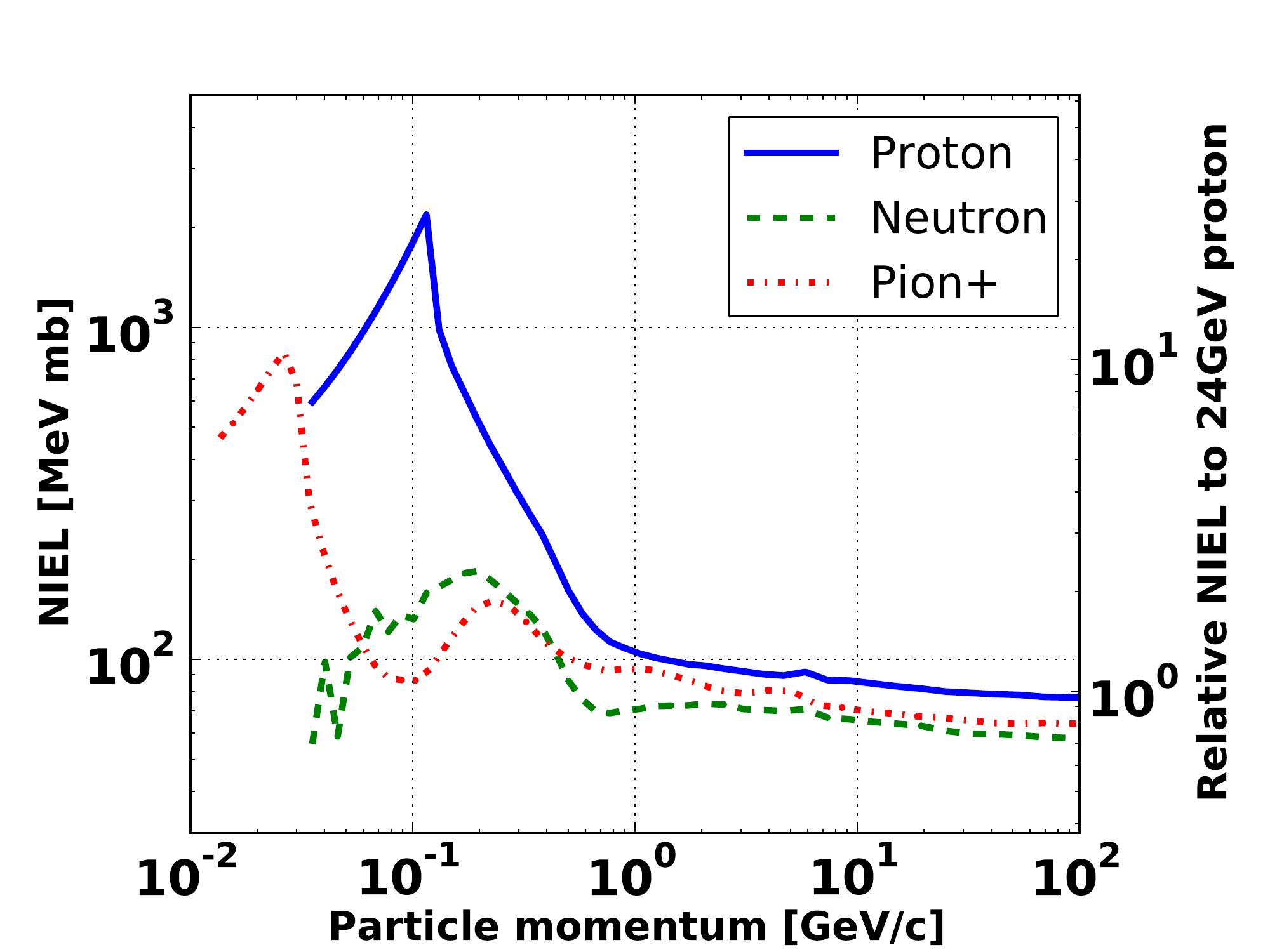}
\label{fig:si_niel_mom}
}
\subfigure[\textcolor{red}{NIEL} of \textcolor{violet}{Diamond}]{
\includegraphics[trim=0cm 0cm 0.5cm 0cm, clip=true,width=0.33\textwidth]{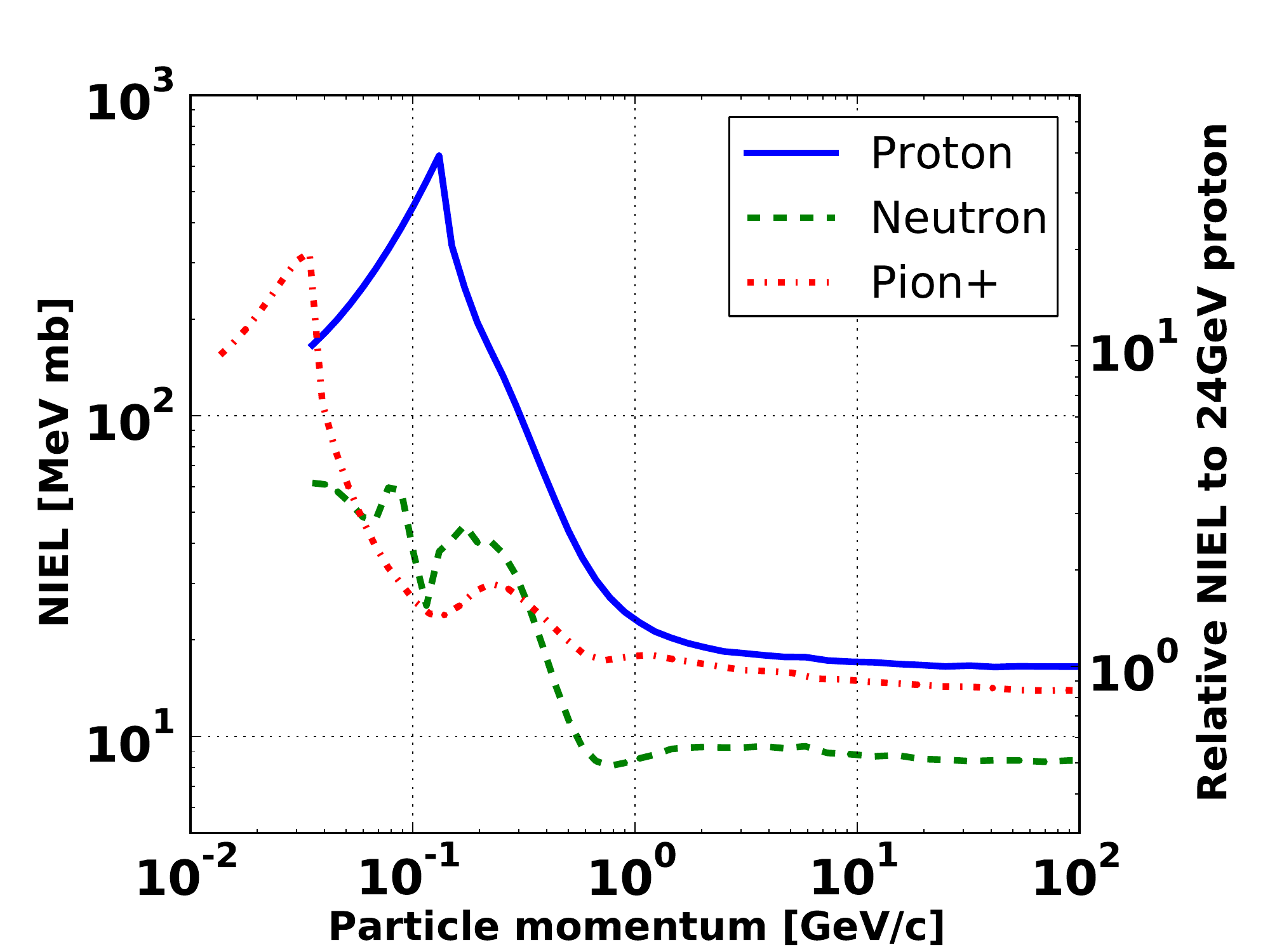}
\label{fig:niel_mom}
}
\subfigure[\textcolor{red}{NIEL} ratio: \textcolor{blue}{Silicon}/\textcolor{violet}{Diamond}]{
\includegraphics[trim=0cm 0cm 0.5cm 0cm, clip=true,width=0.33\textwidth]{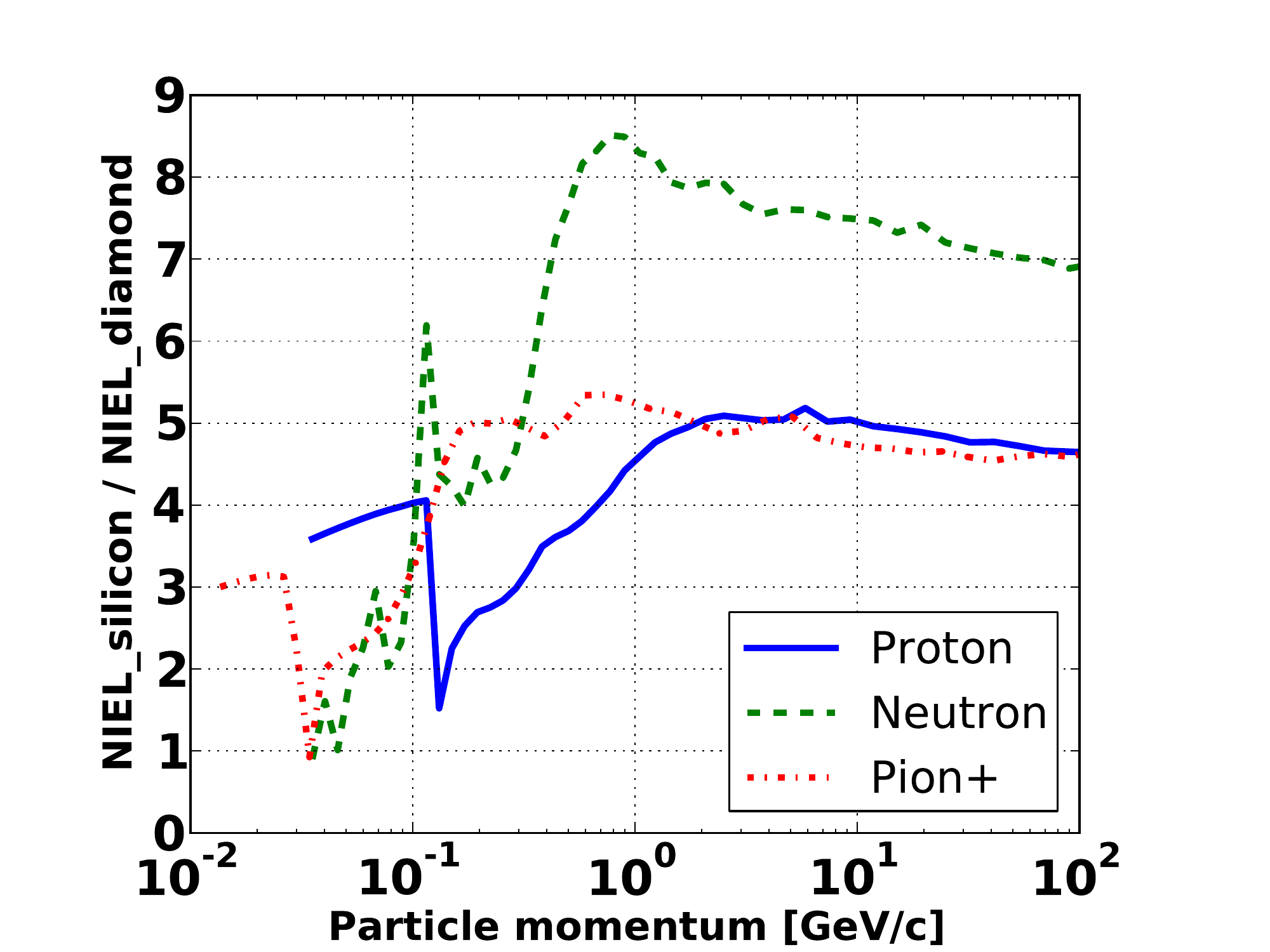}
\label{fig:sivsc_niel_mom}
}
\subfigure[\textcolor{dgreen}{DPA} of \textcolor{blue}{Silicon}]{
\includegraphics[trim=0cm 0cm 0.5cm 0cm, clip=true,width=0.33\textwidth]{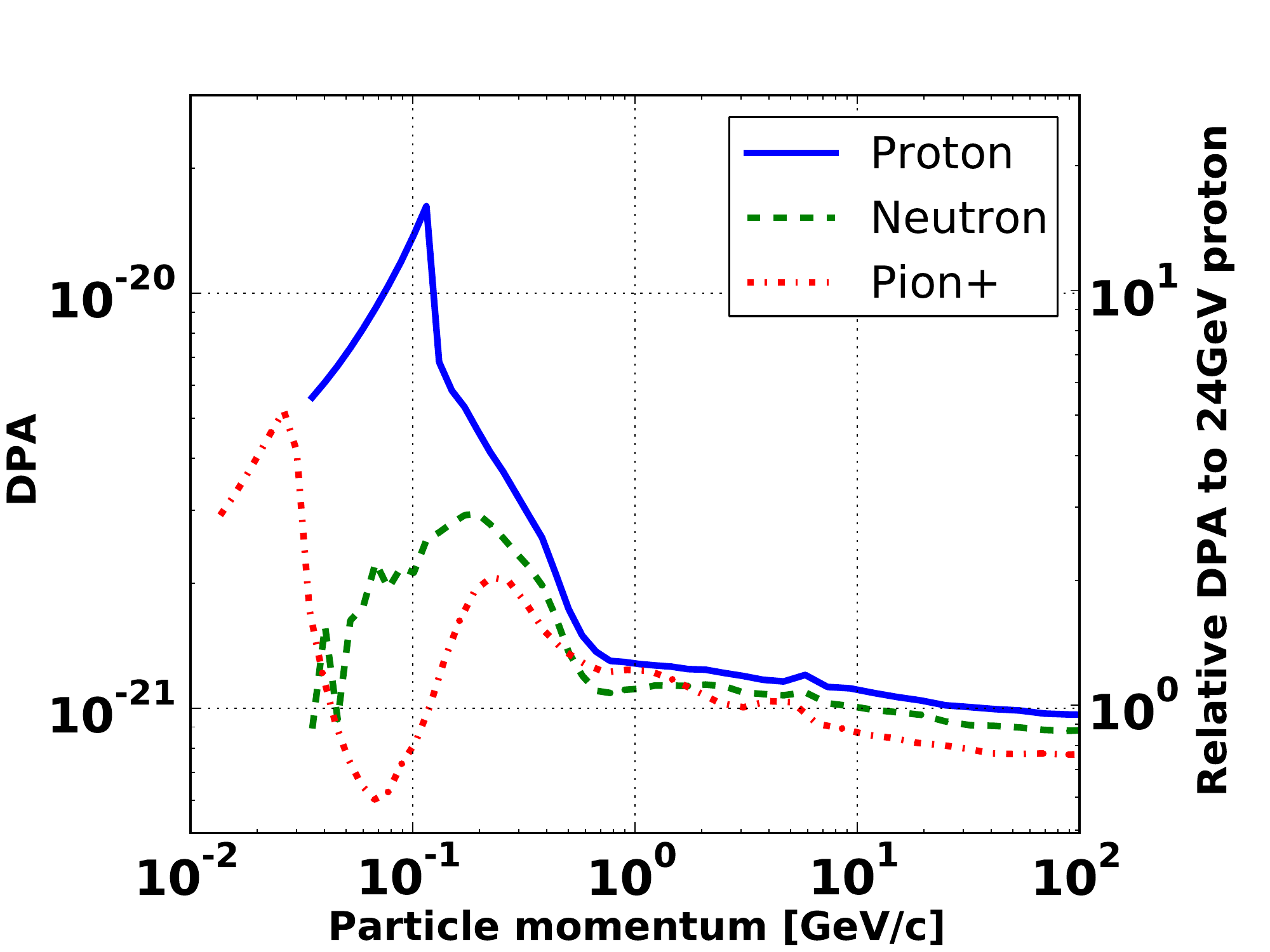}
\label{fig:si_dpa_mom}
}
\subfigure[\textcolor{dgreen}{DPA} of \textcolor{violet}{Diamond}]{
\includegraphics[trim=0cm 0cm 0.5cm 0cm, clip=true,width=0.33\textwidth]{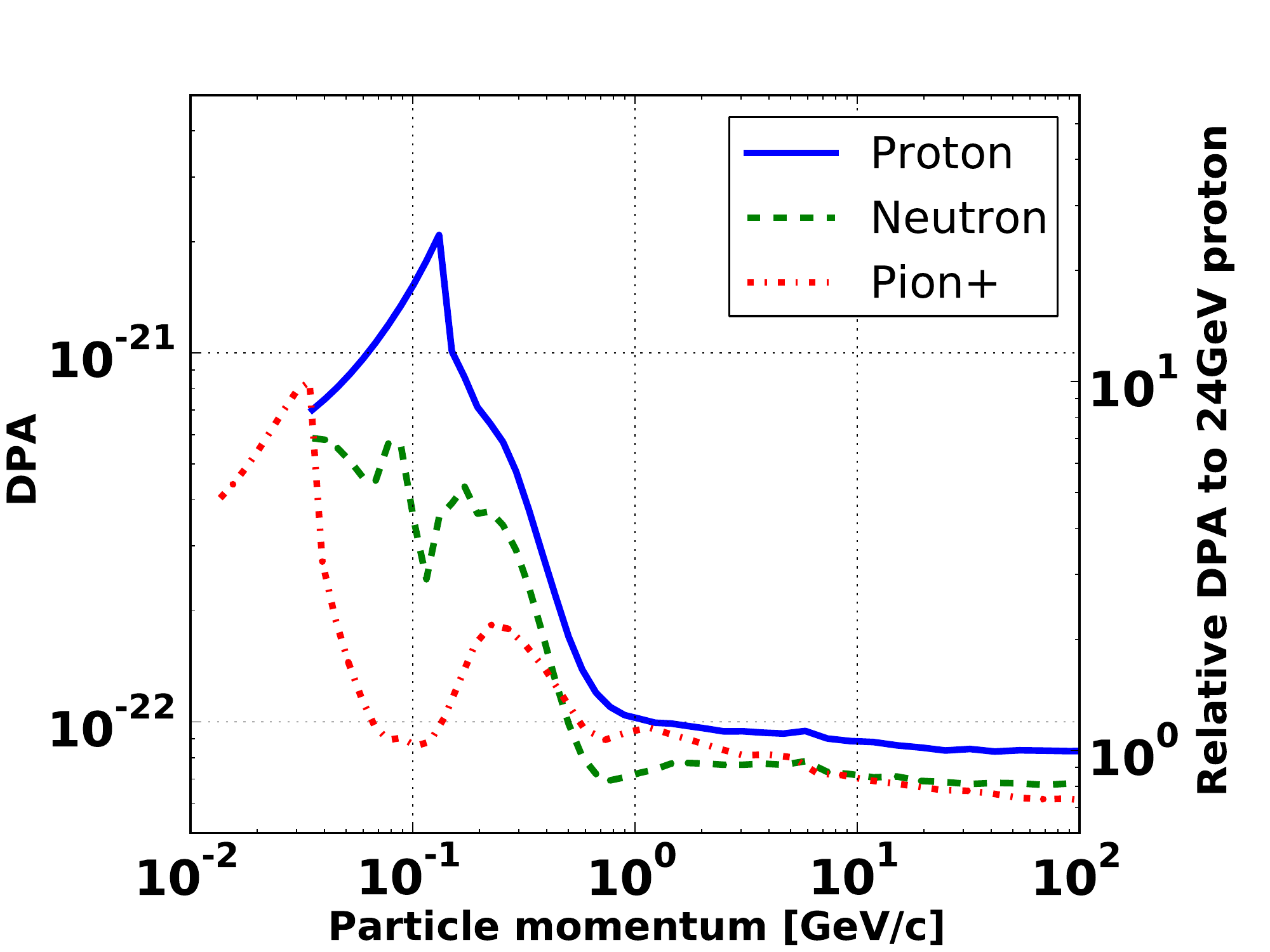}
\label{fig:dpa_mom}
}
\subfigure[\textcolor{dgreen}{DPA} ratio: \textcolor{blue}{Silicon}/\textcolor{violet}{Diamond}]{
\includegraphics[trim=0cm 0cm 0.5cm 0cm, clip=true,width=0.33\textwidth]{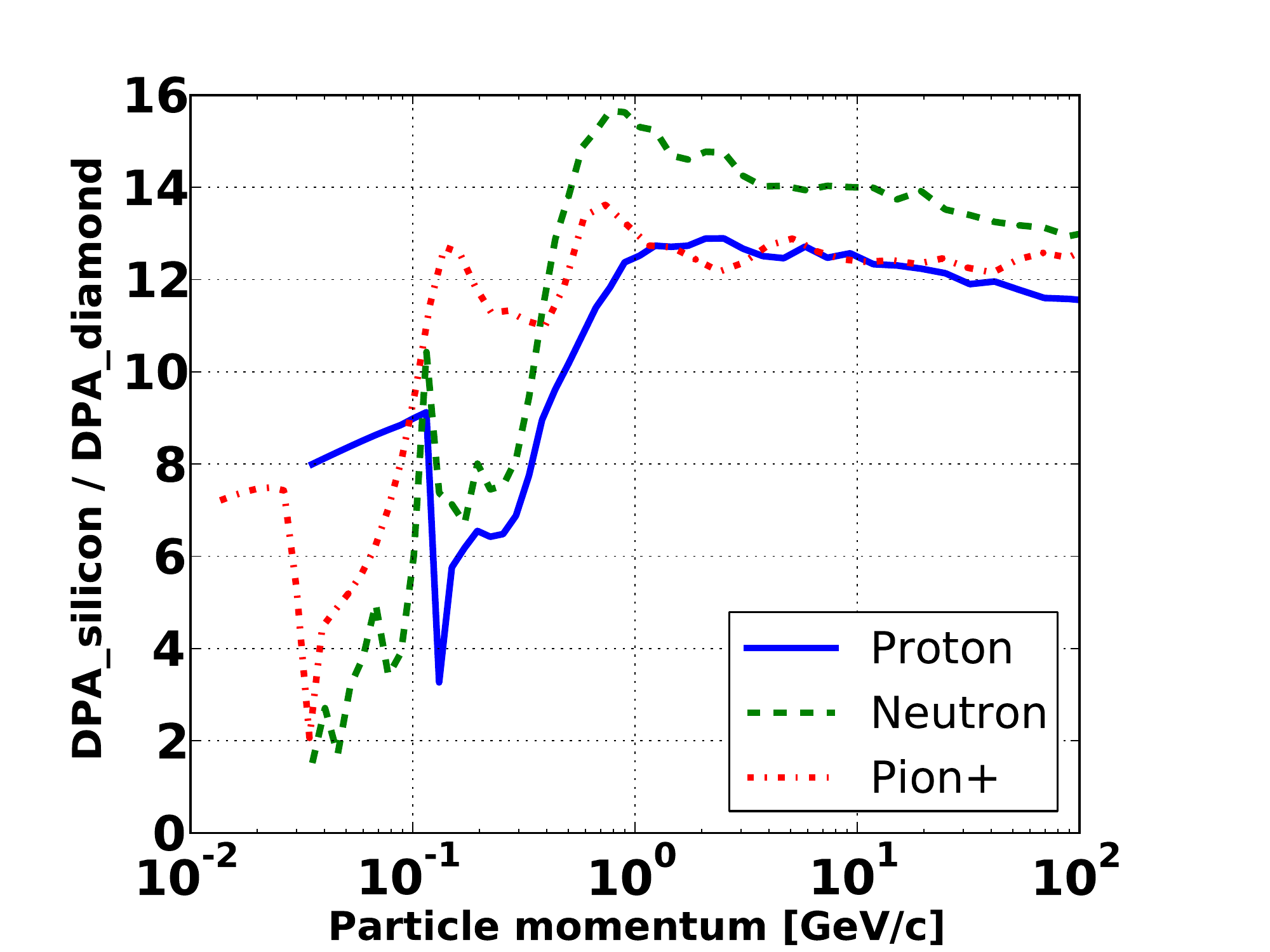}
\label{fig:sivsc_dpa_mom}
}
\subfigure[Ratio: \textcolor{dgreen}{DPA}/\textcolor{red}{NIEL} for \textcolor{blue}{Silicon}]{
\includegraphics[trim=0cm 0cm 0.5cm 0cm, clip=true,width=0.33\textwidth]{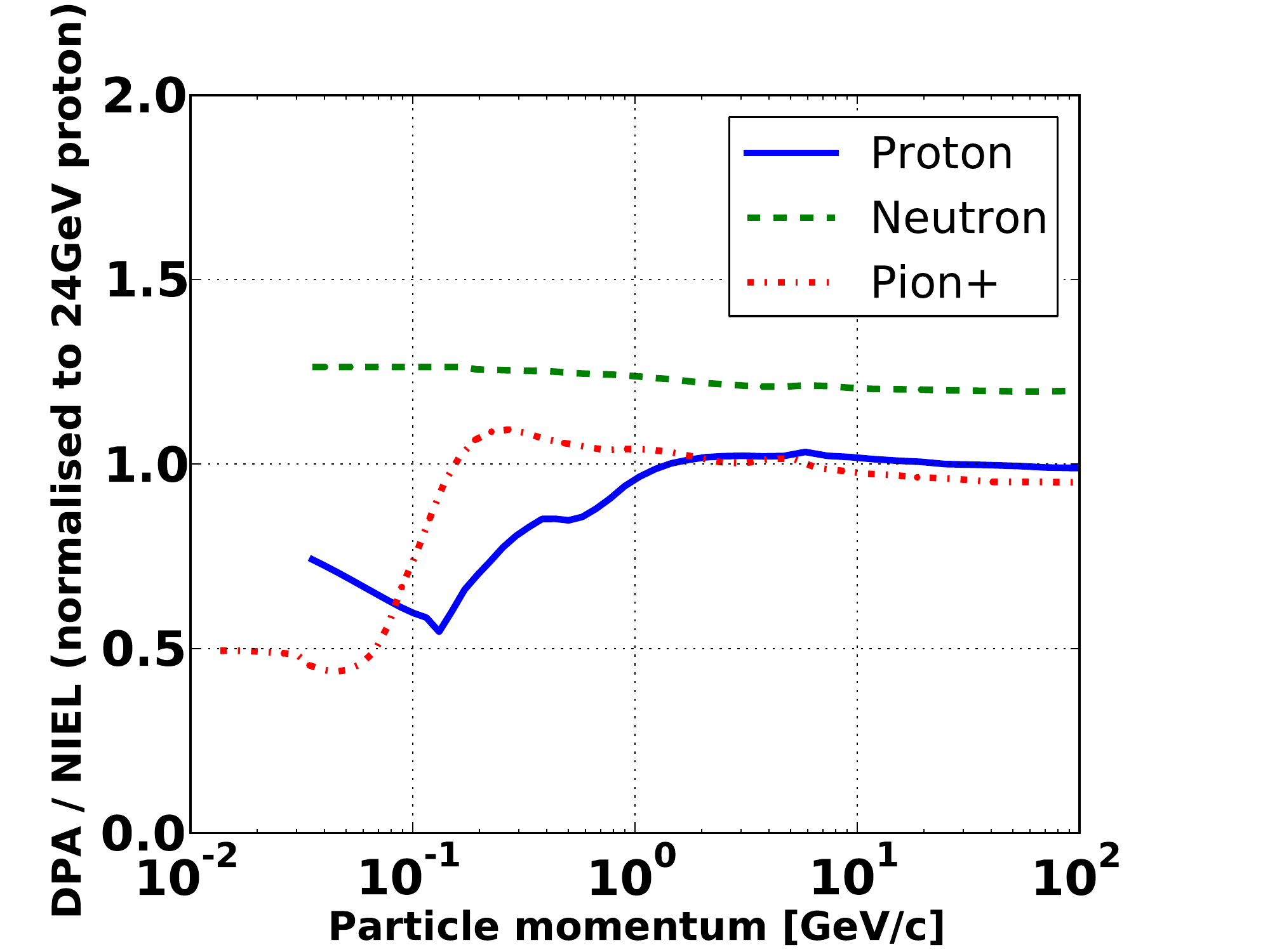}
\label{fig:si_niel_dpa_ratio_mom}
}
\subfigure[Ratio: \textcolor{dgreen}{DPA}/\textcolor{red}{NIEL} for \textcolor{violet}{Diamond}]{
\includegraphics[trim=0cm 0cm 0.5cm 0cm, clip=true,width=0.33\textwidth]{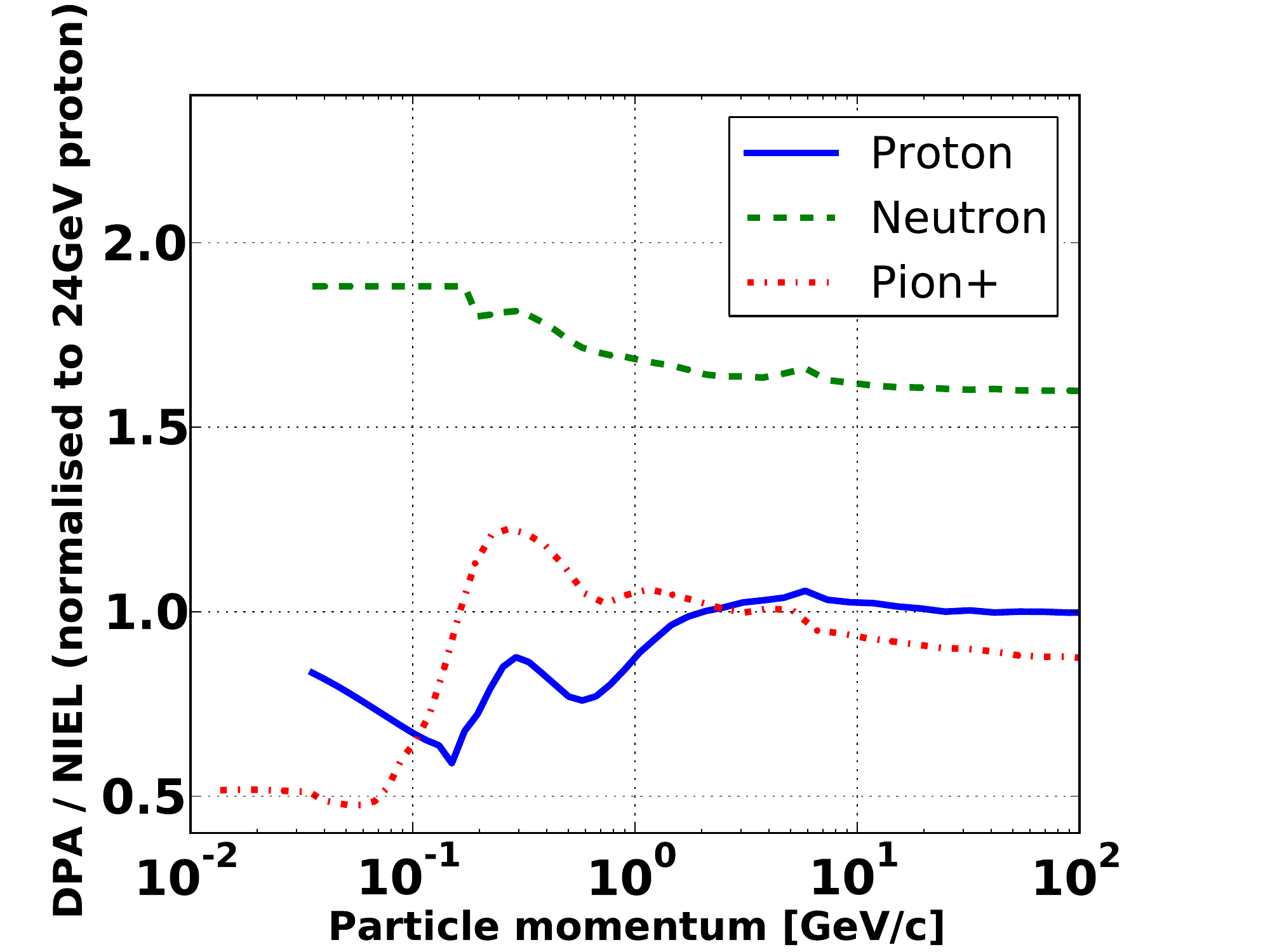}
\label{fig:niel_dpa_ratio_mom}
}
\caption{The simulated results for NIEL and DPA as function of momentum of the impinging particle for silicon and diamond are shown. Fig.~\subref{fig:si_niel_mom} and \subref{fig:niel_mom} show the NIEL results for silicon and diamond and fig.~\subref{fig:sivsc_niel_mom} the ratio of both results. Fig.~\subref{fig:si_dpa_mom} and \subref{fig:dpa_mom} show the DPA results and fig.~\subref{fig:sivsc_niel_mom} the ratio. Fig.~\subref{fig:si_niel_dpa_ratio_mom} and \subref{fig:niel_dpa_ratio_mom} show the ratios of DPA vs. NIEL for the silicon and the diamond case.}

\end{figure*}

In this section the simulation results for protons, neutrons and pions are shown. The quantities that have been determined are the NIEL and the DPA for silicon and diamond. All simulation parameters are the same for silicon and diamond, just the material for the target was changed. All values are normalised per incoming beam particle for the given geometry. 

The energy range of interest for all particles is between 1~MeV and 100~GeV kinetic energy. For a better comparison of particles with different masses all values are shown as function of particle momentum. In the appendix all plots are also shown as function of kinetic energy. All values are given in absolute scale and as a relative number to the corresponding 24~GeV proton value. This value was chosen as reference, as most experimental data exists for 24~GeV protons. This is also a good estimate for all high energetic particles, as all quantities are quite constant for high energies.

The simulated NIEL cross section curves as function of momentum for different particles, are shown in Fig.~\ref{fig:si_niel_mom} for a silicon target and \ref{fig:niel_mom} for a diamond target.
%It is between a factor 100 (for neutrons) and 3000 (for charged particles) smaller than the ionising energy loss.
Figure~\ref{fig:sivsc_niel_mom} shows the ratio between the NIEL curves of silicon and diamond, which directly compares the radiation damage inflicted to silicon or diamond. For high momentum particles (\textgreater\,1\,GeV/c) the diamond receives less damage by factor 5 to 8 (depending also on the particle type). As explained in \cite{boer07} this is due to the fact that the radiation damage is mainly caused by the slow secondary nuclear fragments, which are more numerous in case of the heavier silicon nucleus. At lower momentum the difference in radiation damage is smaller.

Not all NIEL leads to material damage. As long the recoil energy is below the lattice damage threshold, only non-damaging phonon interactions are taking place. 
The NIEL, which produces recoils above the lattice displacement energy, is used to calculate the number of displaced atoms. The DPA curves are shown in Fig.~\ref{fig:si_dpa_mom} for silicon, \ref{fig:dpa_mom} for diamond and \ref{fig:sivsc_dpa_mom} shows the ratio between silicon and diamond. The ratio plot shows that in diamond the number of displacements produced is about a factor 12 to 14 less for high momentum particles, a higher difference than in the NIEL scaling. Again at low momentum the difference is smaller. Well below 100\,MeV/c the ratio should converge to 3.6, which corresponds to the ratio of the cross section of the Rutherford scattering \cite{boer07}. This can't be shown with these simulations since the particles are completely stopped by ionizing energy deposition and the curve jumps to a value that does not correspond to the actual DPA curve.

For the charged particles at low momentum ($\sim$100\,MeV/c for protons and $\sim$30\,MeV/c for pions) the curve sharply decreases. At this range of momentum the particle is fully stopped inside the diamond and the total kinetic energy is deposited. The position of the kink depends on the thickness of the detector.

Figures~\ref{fig:si_niel_dpa_ratio_mom}~and~\ref{fig:niel_dpa_ratio_mom} show the ratio between DPA and NIEL for the silicon and for the diamond case. The curves are normalised to the value for 24\,GeV protons and hence the high energetic protons have a ratio of one.
High energetic pions show the same ratio as the protons, but neutrons lead to more damage in the DPA scaling. For silicon (diamond) this effect is about 20\,\% (60\,-\,90\,\%).
At lower momentum more phonon interactions take place. Since phonon interactions are excluded in the DPA calculation, but are included in the NIEL, pions and protons at lower momentum (below 100\,MeV/c) produce less radiation damage in the DPA than obtained with the NIEL calculation (ratio \textless\,1).

%Also due to phonon interactions included in the NIEL, but not in the DPA, the relative values for pions and protons at lower momentum (below 100\,MeV/c) have less weight in the DPA scaling. 

\section{Simulations of a 1\,$\mu$m thick detector}
\label{sec:thin_detector}
The results of simulations of a 400\,$\mu$m thick detector presented in section \ref{sec:results} are not representative for charged primary particles in the low energetic region since they are fully stopped by ionization and the damage curve has a downwards kink towards lower energies. To simulate the radiation damage parameter for lower energies the same simulation was performed for a 1\,$\mu$m thick detector. The NIEL result of this simulation is used for the comparison with the SRIM data shown in fig.~\ref{fig:niel_srimvsfluka}.

\begin{figure}[t]
\begin{center}
\includegraphics[width=0.8\linewidth]{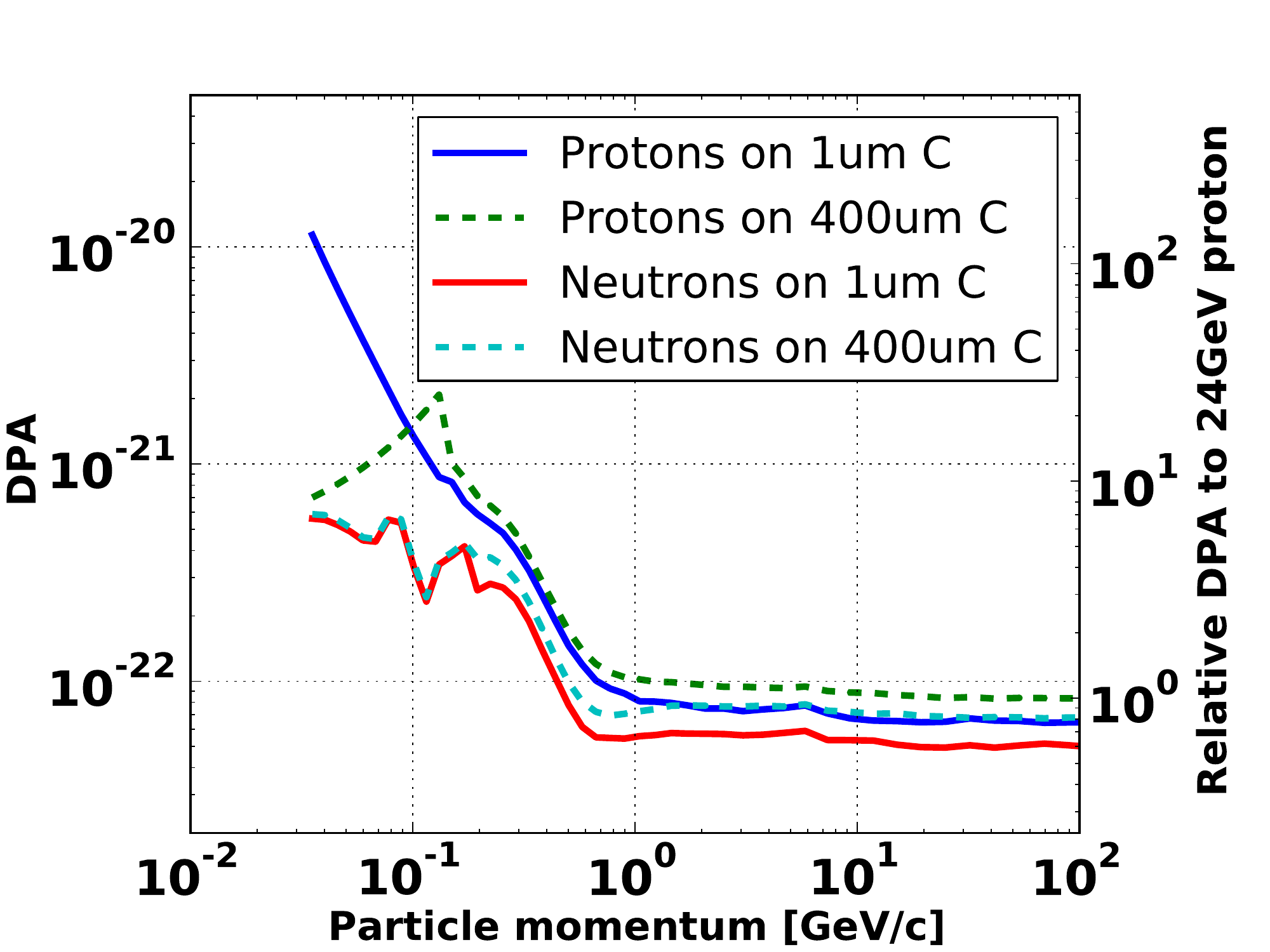}
\end{center}
\caption{Generated displacements per atom by indicated particles at a given momentum in a 1\,$\mu$m thick piece of diamond in comparison with a 400\,$\mu$m thick piece of diamond.}
\label{fig:dpa_thin}
\end{figure}

The DPA results for diamond detectors of different thicknesses are shown in figure \ref{fig:dpa_thin}. The downward kink at low energies in the proton curve is not present in the 1\,$\mu$m data. The neutron data for low energies matches the data from the thick detector. At higher energies the data for a thin detector is below the data for the thick detector. Secondary particles can produce additional displacements in the thick detector, but in the thin detector they escape the detector and fewer displacements are produced.

\section{Impact to real detector systems}
The amount of traps in the bulk material is a key parameter for the quality of the diamond. The signal output of a real life detector depends not only on the CCD as function of the trap density as described in chapter \ref{sec:detect_efficiency}.
A more important aspect is the rate dependence of the charge trapping. At high incident particle rates the traps are quickly filled and become inactive. In this case the CCD and with this the signal first increases.
This process is often referred to pumping or priming. However, under the influence of the bias voltage more electrons (holes) are trapped at the positive (negative) electrode, which leads to an internal electric field opposite to the bias voltage (polarization field). 
It was shown that this polarization effect can squeeze the electric field out of the diamond and thereby reduce the effective detector thickness \cite{pom2008}. Polarization effects can dominate the decrease of the signal strength and kick in at much lower fluences than the effects of reduced CCD\cite{guthoff2013}. The polarisation increases at high rates of incident particles, but the effect of its corresponding internal electric field can be reduced if the external bias voltage is increased. This makes a comparison of data from different test beams difficult if the measurements have not been performed with the same detector setup and the same intensity.

\section{Summary and outlook}
A novel way of simulating the radiation damage effects in solid state detectors, using the displacement of atoms calculation of the FLUKA simulations package, has been presented. This model takes more effects into account than the NIEL traditionally used and is therefore more suitable as radiation damage scaling rule. The obtained plots of proton, neutron and pion irradiation of diamond and silicon material at a wide range of energies can be used for comparisons of radiation damage studies at different energies and for different particles. In case of a diamond detector it is important to eliminate polarisation effects, if one wants to compare with DPA calculations, e.g. by doing measurements at high bias voltages and/or low intensities. In real experiments with high fluxes the signal decrease is typically much faster than predicted by the DPA calculations, especially in environments near hadronic calorimeters, where the flux of low energy neutrons is high\cite{guthoff2013}.

\section{Acknowledgements}
We gratefully acknowledge the financial support from the Bundesministerium f\"ur Bildung und Forschung (BMBF) in Germany
and the members of the CMS  Beam Radiation Monitoring group for useful discussions.

\bibliography{diss}

%\pagebreak
\begin{appendix}
\section{Plots as function of kinetic energy}

\begin{figure*}[p]
\subfigure[\textcolor{red}{NIEL} of \textcolor{blue}{Silicon}]{
\includegraphics[trim=0cm 0cm 0.5cm 0cm, clip=true,width=0.33\textwidth]{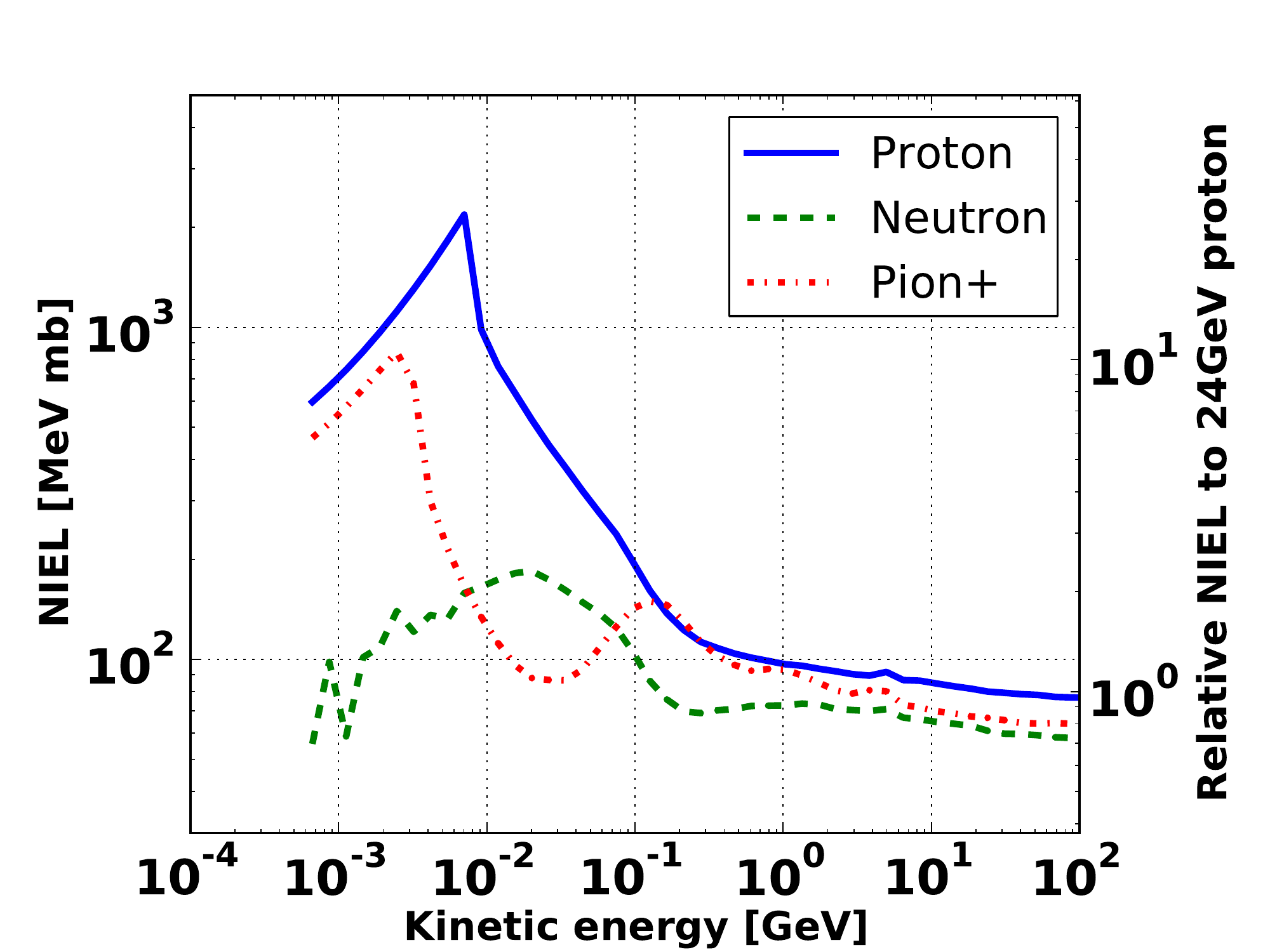}
\label{fig:si_niel_e}
}
\subfigure[\textcolor{red}{NIEL} of \textcolor{violet}{Diamond}]{
\includegraphics[trim=0cm 0cm 0.5cm 0cm, clip=true,width=0.33\textwidth]{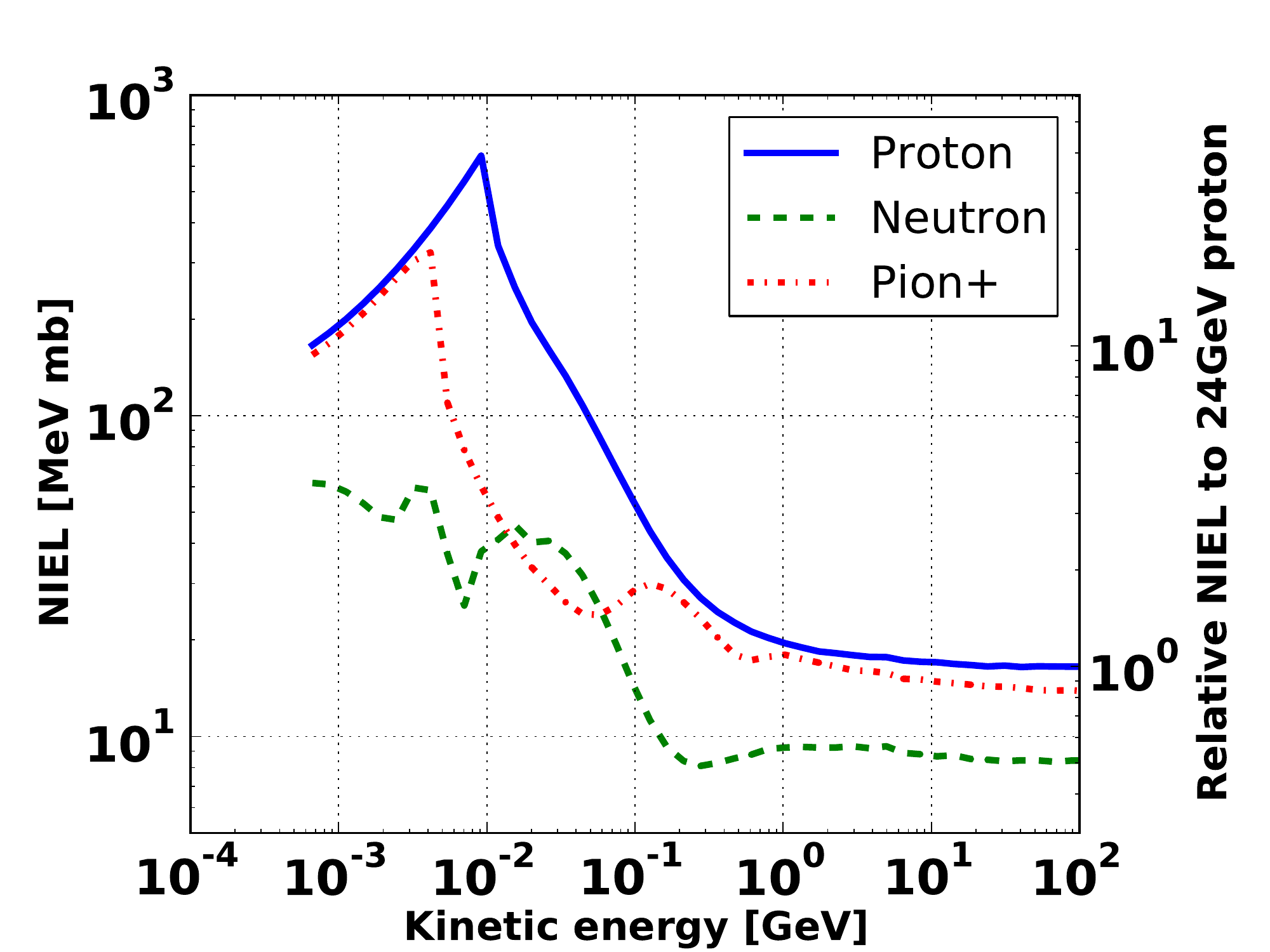}
\label{fig:niel_e}
}
\subfigure[\textcolor{red}{NIEL} ratio: \textcolor{blue}{Silicon}/\textcolor{violet}{Diamond}]{
\includegraphics[trim=0cm 0cm 0.5cm 0cm, clip=true,width=0.33\textwidth]{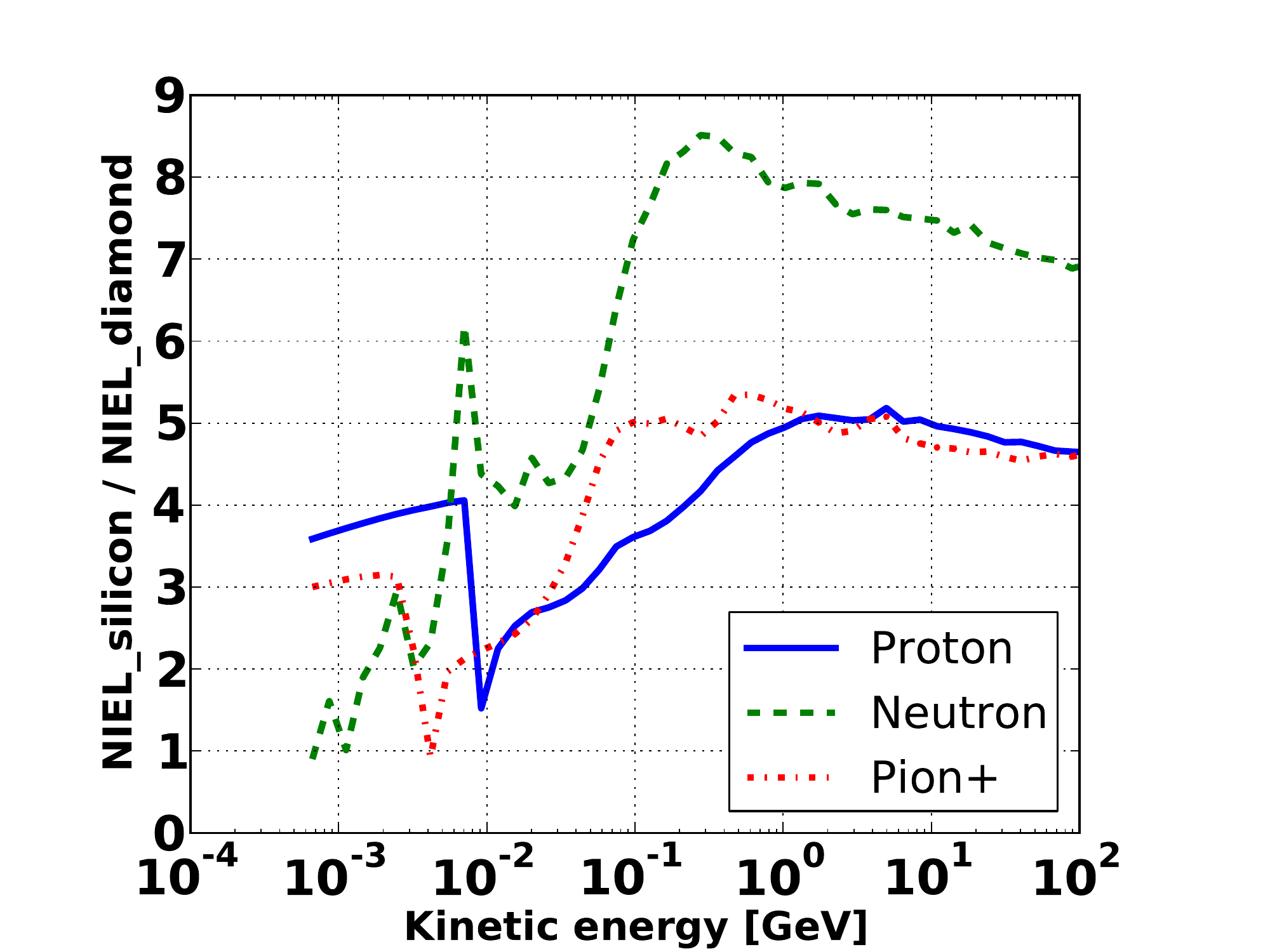}
\label{fig:sivsc_niel_e}
}
\subfigure[\textcolor{dgreen}{DPA} of \textcolor{blue}{Silicon}]{
\includegraphics[trim=0cm 0cm 0.5cm 0cm, clip=true,width=0.33\textwidth]{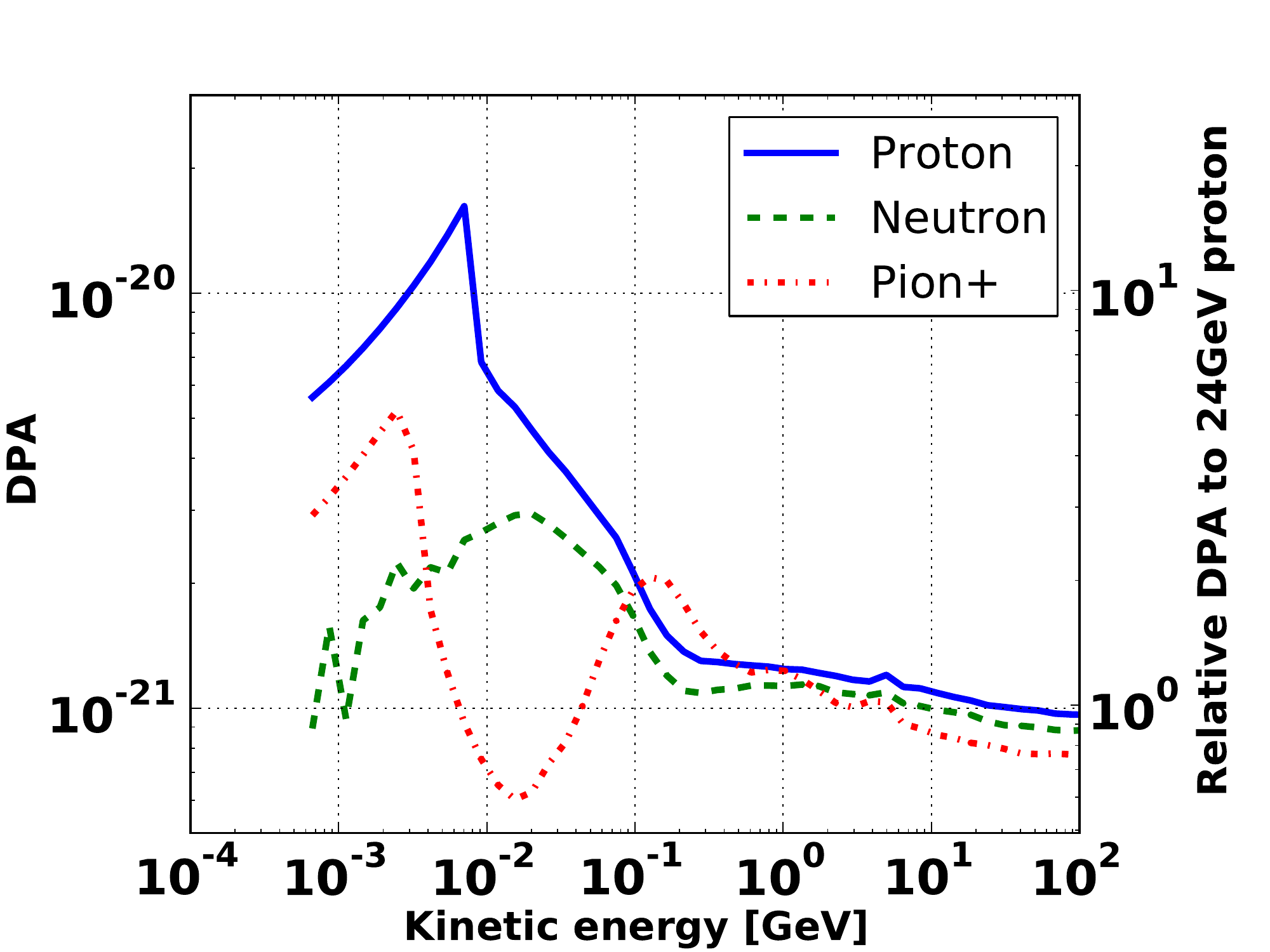}
\label{fig:si_dpa_e}
}
\subfigure[\textcolor{dgreen}{DPA} of \textcolor{violet}{Diamond}]{
\includegraphics[trim=0cm 0cm 0.5cm 0cm, clip=true,width=0.33\textwidth]{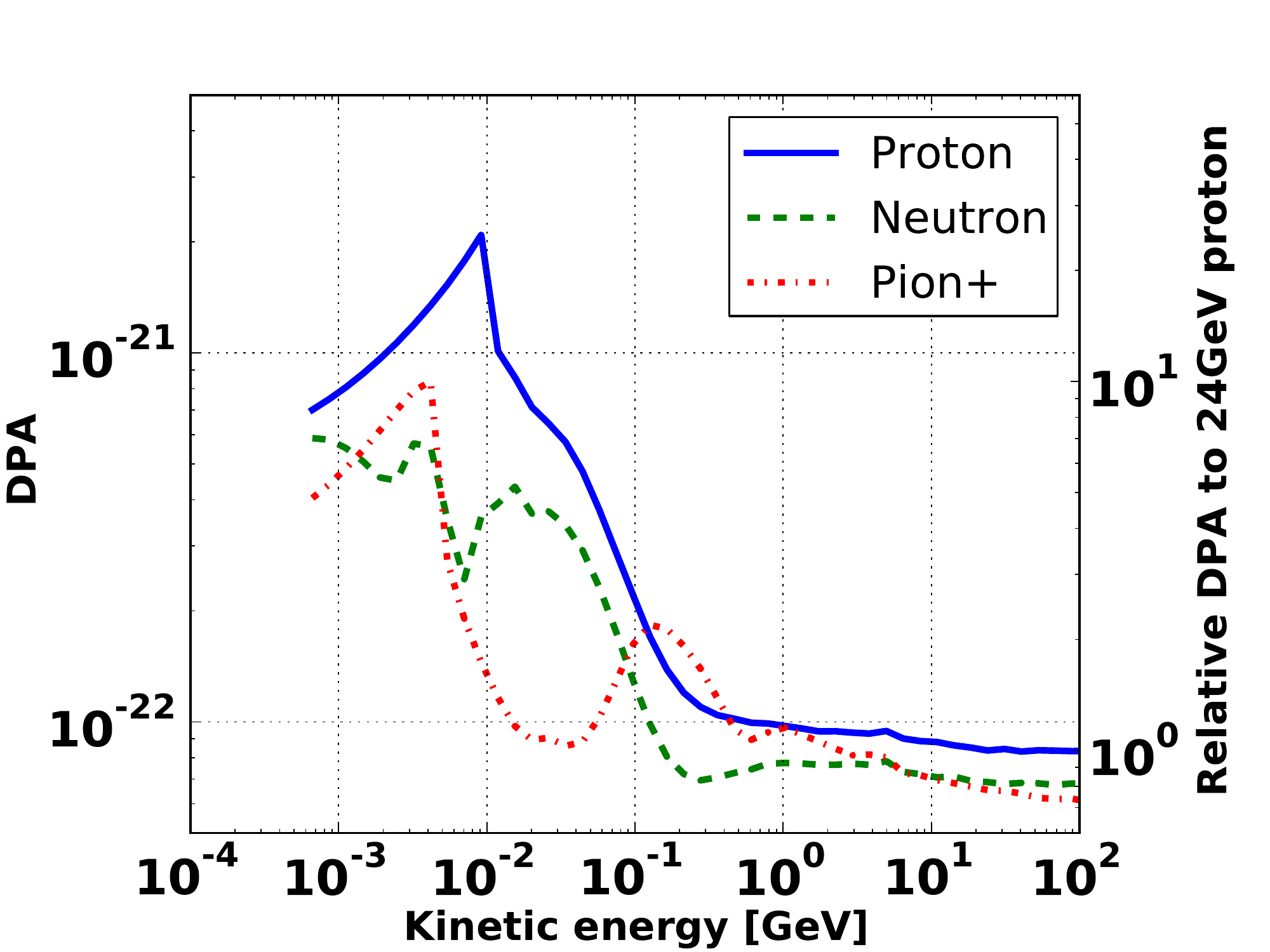}
\label{fig:dpa_e}
}
\subfigure[\textcolor{dgreen}{DPA} ratio: \textcolor{blue}{Silicon}/\textcolor{violet}{Diamond}]{
\includegraphics[trim=0cm 0cm 0.5cm 0cm, clip=true,width=0.33\textwidth]{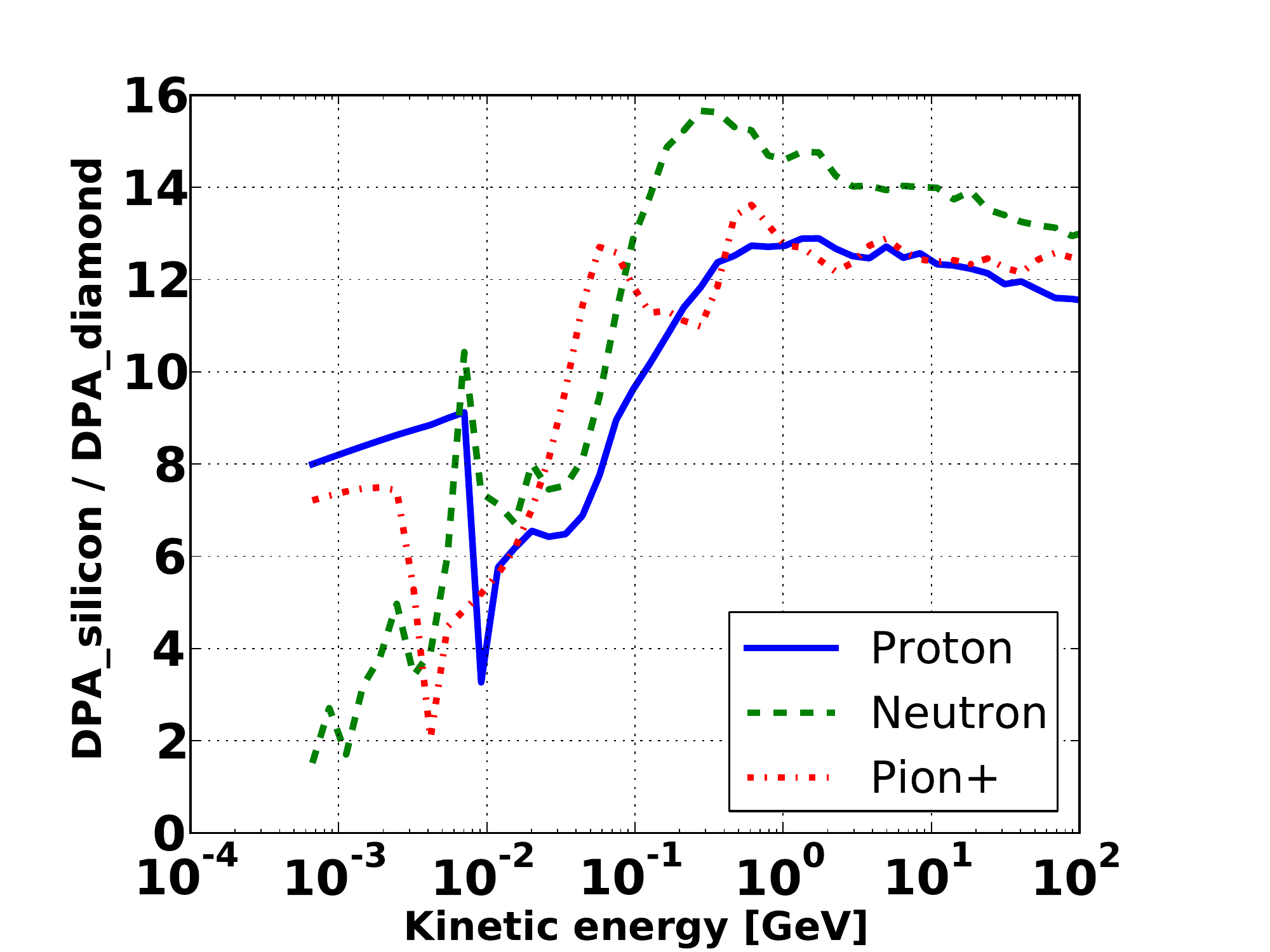}
\label{fig:ratio_si_dia_dpa_e}
}
\subfigure[Ratio: \textcolor{dgreen}{DPA}/\textcolor{red}{NIEL} for \textcolor{blue}{Silicon}]{
\includegraphics[trim=0cm 0cm 0.5cm 0cm, clip=true,width=0.33\textwidth]{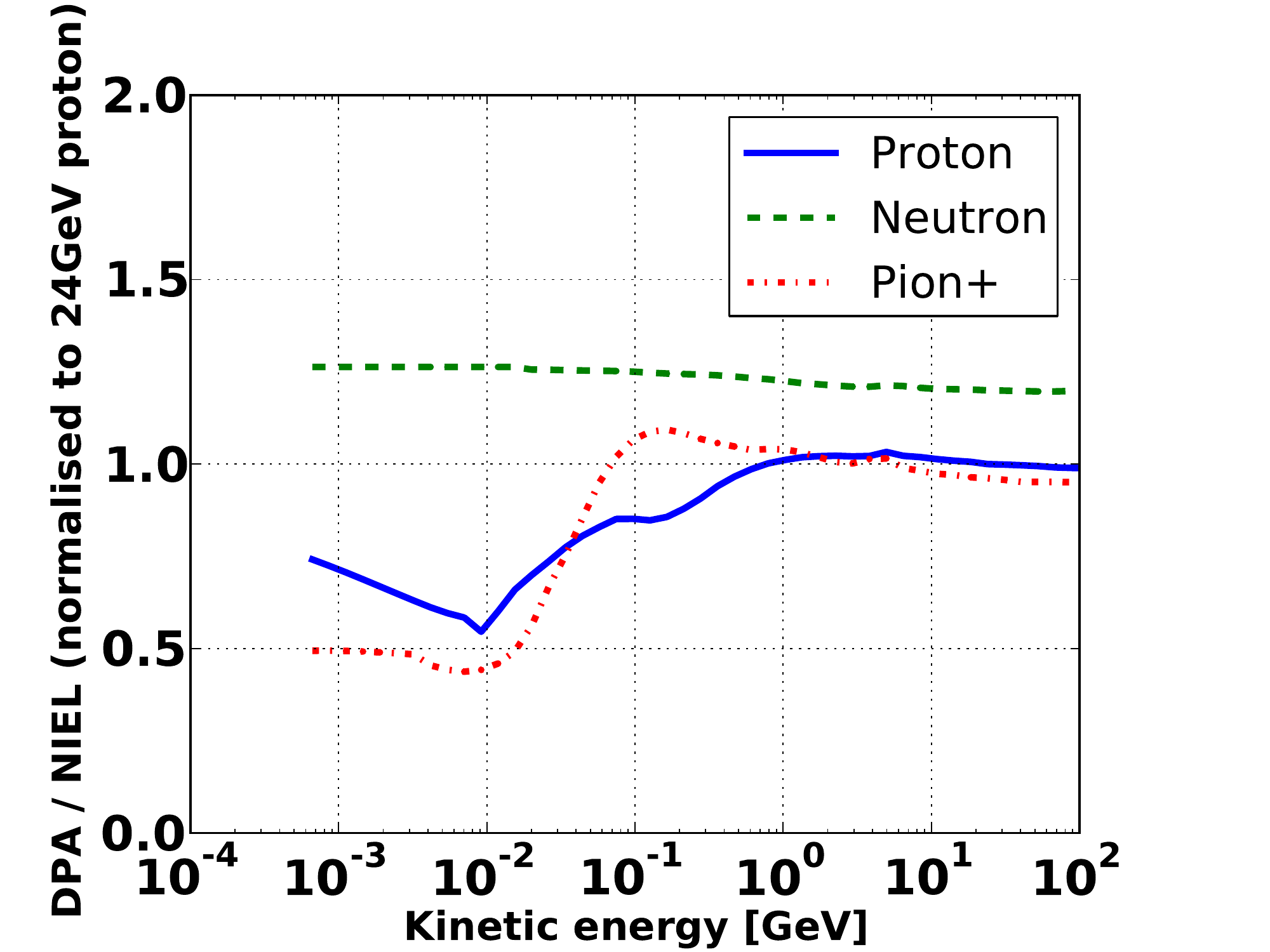}
\label{fig:si_niel_dpa_ratio_e}
}
\subfigure[Ratio: \textcolor{dgreen}{DPA}/\textcolor{red}{NIEL} for \textcolor{violet}{Diamond}]{
\includegraphics[trim=0cm 0cm 0.5cm 0cm, clip=true,width=0.33\textwidth]{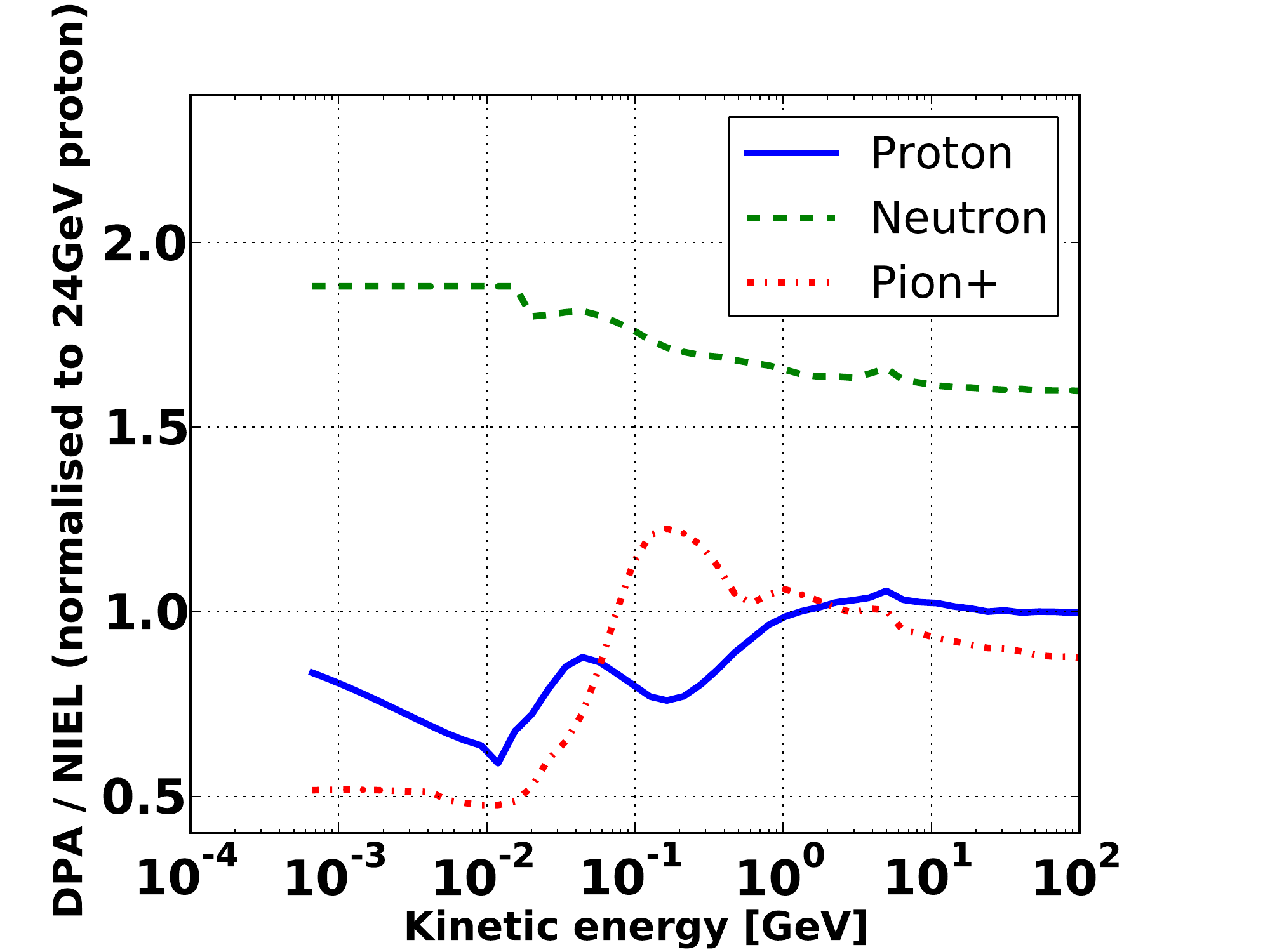}
\label{fig:niel_dpa_ratio_e}
}
\caption{The simulated results for NIEL and DPA as function of kinetic energy of the impinging particle for silicon and diamond are shown. Fig.~\subref{fig:si_niel_e} and \subref{fig:niel_e} show the NIEL results for silicon and diamond and fig.~\subref{fig:sivsc_niel_e} the ratio of both results. Fig.~\subref{fig:si_dpa_e} and \subref{fig:dpa_e} show the DPA results and fig.~\subref{fig:sivsc_niel_e} the ratio. Fig.~\subref{fig:si_niel_dpa_ratio_e} and \subref{fig:niel_dpa_ratio_e} show the ratios of DPA vs. NIEL for the silicon and the diamond case.}
\end{figure*}

\newpage

\begin{figure}[p]
\begin{center}
\includegraphics[width=0.7\linewidth]{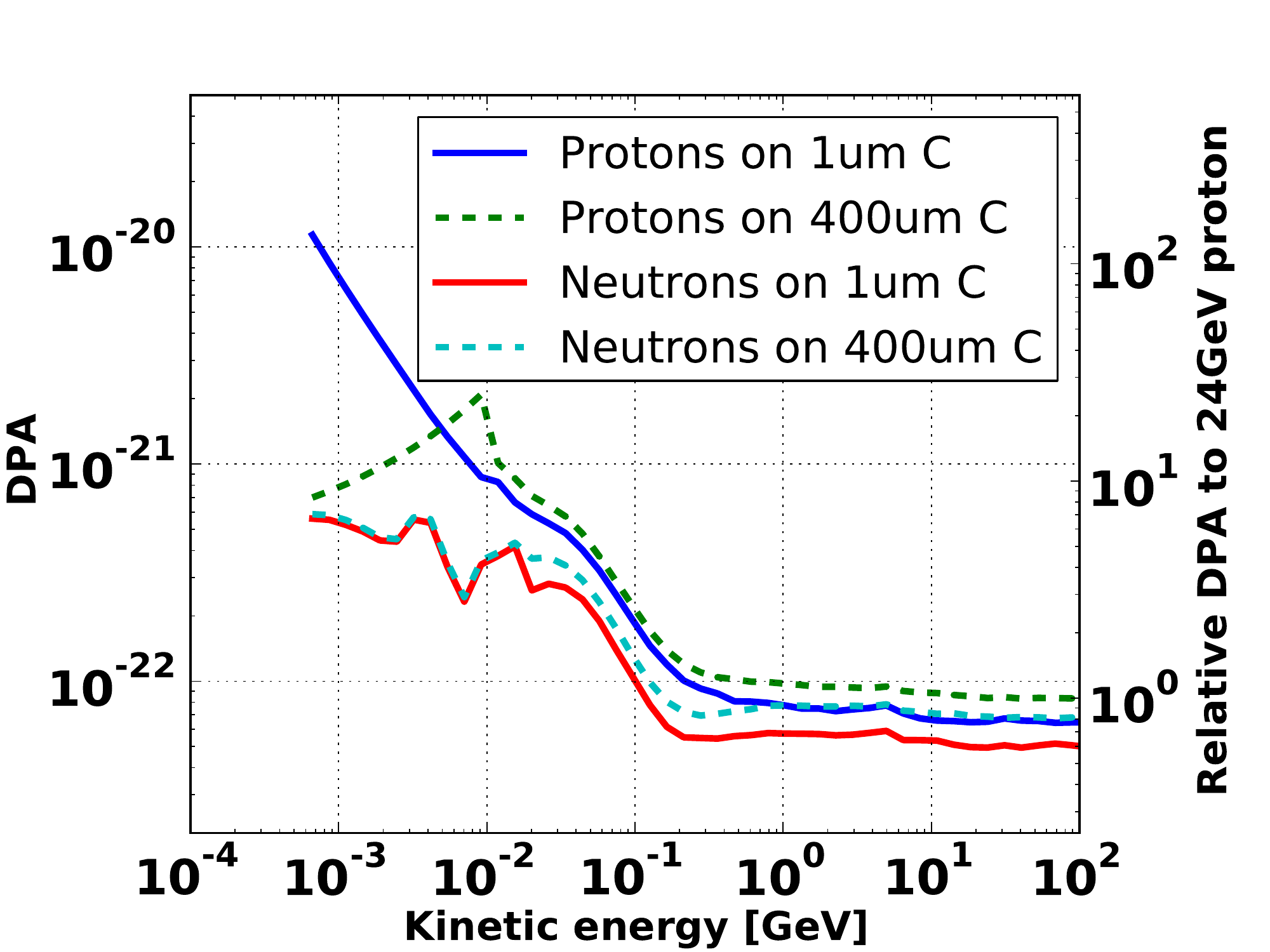}
\end{center}
\caption{Generated displacements per atom by indicated particles at a given energy in a 1\,$\mu$m thick piece of diamond in comparison with a 400\,$\mu$m thick piece of diamond.}
\label{fig:dpa_thin_e}
\end{figure}

\end{appendix}

\end{document}